\documentclass[twocolumn,showpacs,preprintnumbers,amsmath,amssymb]{revtex4}
\usepackage{amssymb}
\usepackage{txfonts}

\usepackage{graphicx}
\usepackage{dcolumn}
\usepackage{bm}

\begin{document}

\title{Entanglement and quantum discord dynamics of two atoms under practical feedback control}

\author{Yang Li}
\author{Bin Luo}
\author{Hong Guo}\thanks{Author to whom correspondence should be
addressed. phone: +86-10-6275-7035, Fax: +86-10-6275-3208, E-mail:
hongguo@pku.edu.cn.} \affiliation{CREAM Group, State Key Laboratory
of Advanced  Optical Communication Systems and Networks, School of
Electronics Engineering and Computer Science, and Center for
Computational Science and Engineering (CCSE), Peking University,
Beijing 100871, P. R. China}%

\date{\today}

\begin{abstract}
We study the dynamics of two identical
atoms resonantly coupled to a single-mode cavity under practical feedback
control, and focus on the detection inefficiency. The entanglement is
induced to vanish in finite time by the inefficiency of detection.
Counterintuitively, the asymptotic entanglement and quantum discord can be
increased by the inefficiency of detection. The noise of detection triggers control field to create
entanglement and discord when no photon are emitted from the atoms. Furthermore, sudden change happens to the dynamics of entanglement.
\end{abstract}

.

\maketitle

\section{Introduction}
Correlations are crucial to information science. There
have been a lot of studies on entanglement, a special quantum
correlation, because many quantum information processes depend on
entanglement \cite{QI}. Recently, new studies show that
separable states can speed up some computational task compared to
classical computation \cite{speedup}. A more general quantum correlation, quantum
discord, has also received a great deal of attention \cite{discord1,discord2,discord3,discord4,pictorial,almost,Xstate}.

The crucial quantum properties can be destroyed by
the influence of the environment. Many studies were done on this subject \cite{ESD,revival,Entanglementdynamics,suddenchange1,suddenchange2,correlationdynamics}.
And many interesting phenomena have been found, such as
entanglement sudden death \cite{ESD}, entanglement revival \cite{revival}, and sudden change
for quantum discord \cite{suddenchange1,suddenchange2}. In order to avoid or delay the influence of the
environment, many methods have been proposed, such as quantum error
correction \cite{errorcorrection1,errorcorrection2}, decoherence-free
subspace \cite{DFS1,DFS2,DFS3}, and dynamical control
\cite{Dynamicalcomtrol1,Dynamicalcomtrol2,Dynamicalcomtrol3}.
Among them, quantum feedback control \cite{feedback} is believed to be a promising method.
For a system consisting of two two-level atoms coupled to a single-mode cavity which
is heavily damped, the entanglement of the steady state can be improved by
Markovian feedback control\cite{model,feedbackentanglement0}.
Effects of different feedback Hamiltonians and detection processes on entanglement generation were
explored \cite{feedbackentanglement1,feedbackentanglement2,feedbackentanglement3}. In
these works, the detection is assumed to be perfect, but it is very
hard to achieve in practice. The influence of the inefficiency of
detection on the dynamical and asymptotic behavior remains an
open question.

In the present work, we investigate the dynamics of two identical
atoms resonantly coupled to a single-mode cavity under feedback
control. It was thought that the inefficiency decreases the steady entanglement,
but our study shows that it may also increase the steady entanglement for some initial state.
For the dynamical behavior, entanglement vanishes in finite time due to the inefficiency of
detection. We also find that the noise of detection is an source to trigger the creation of
entanglement and quantum discord. More importantly, we find that
the dynamics of the entanglement may also undergo sudden change.

\section{Theoretical framework}

\begin{figure}
\centering
\includegraphics[height=0.25\textheight,width=0.4\textwidth]{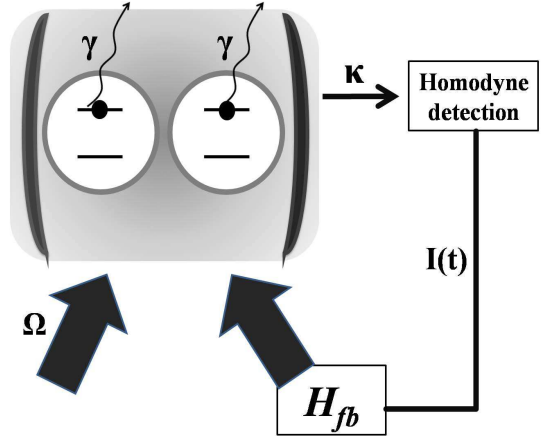}
\caption{\small Schematic view of the model. Two atoms are driven by a laser with Rabi frequency $\Omega$,
and coupled to a heavily damped cavity. The feedback Hamiltonian is applied to the atoms
according to the homodyne current $I(t)$ derived from the leaky cavity.} \label{fig1}
\end{figure}

We investigate the dynamics of two identical atoms resonantly coupled to a single mode
cavity which is driven by a laser field with Rabi frequency $\Omega$ and damped with
decay rate $\kappa$ (See FIG. 1). The atoms couple the cavity with strength $g$ and spontaneously decay
with rate $\gamma$. When the cavity mode is heavily damped, it can be adiabatically
eliminated. And in the limit $\Gamma=g^2 /\kappa \gg \gamma$, the dynamical evolution of this system is described by the Dicke model \cite{feedbackentanglement0,feedbackentanglement1,feedbackentanglement2,feedbackentanglement3}:
\begin{equation}
\frac{{d\rho }}{{dt}} =  - i[H,\rho ] + D[A]\rho ,
\end{equation}
where $\rho$ is the density matrix of the two atoms, $H = \Omega
(\sigma _x^{(1)}  + \sigma _x^{(2)} )$ represents the driving of the
laser, and $D[A]\rho  = A\rho A^\dag   - (A^\dag  A\rho  + \rho
A^\dag  A)/2$ represents the irreversible evolution induced by the
interaction between the system and the environment with the jump
operator $A = \Gamma (\sigma _ - ^{(1)}  + \sigma _ - ^{(2)} ) $.
Without losing generality, we let $\Gamma = 1$.

In this paper, we will consider the Markovian feedback \cite{feedback}, with the control
Hamiltonian $H_{fb} = I(t) F$, where $I(t)$
is the signal from the homodyne detection of the cavity output.

In the homodyne-based scheme, the detector registers a continuous
photocurrent, and the feedback Hamiltonian is constantly applied to
the system. The master equation becomes \cite{feedbackentanglement2,feedbackentanglement3}:
\begin{equation}
\frac{{d\rho }}{{dt}} =  - i[H + \frac{1}{2}(A^\dag  F + FA),\rho ] + D[A - iF]\rho,
\end{equation}
where $F$ is the feedback Hamiltonian. In this paper, we consider
the symmetric feedback Hamiltonian $F =  - \lambda [\mu (\sigma
_x^{(1)} \sigma _z^{(2)}  + \sigma _z^{(1)} \sigma _x^{(2)} ) +
(\sigma _x^{(1)}  + \sigma _x^{(2)} )]$ as in
\cite{feedbackentanglement2}. It reduces to the feedback Hamiltonian
in \cite{feedbackentanglement0} when $\mu =0$.

Practically, the efficiency of the detection, denoted by $\eta$, is less than $1$. The modified master
equation takes the form \cite{feedbackentanglementcondition}:
\begin{equation}
\frac{{d\rho }}{{dt}} =  - i[H + \frac{1}{2}(A^\dag  F + FA),\rho ] + D[c - iF]\rho  + D[\sqrt {\frac{{1 - \eta }}{\eta }} F]\rho.
\end{equation}

In this paper, we choose $\Omega =0$, $\lambda =1$ and
$\mu=1$. For these parameters, the steady state is maximally entangled state $\phi_{+}=(\left| {eg} \right\rangle +\left| {ge}
\right\rangle)/ \sqrt{2}$ if the initial state is in symmetric
subspace for perfect detection \cite{feedbackentanglement2}. In the basis $\{ \left| {gg}
\right\rangle ,\left| {ge} \right\rangle ,\left| {eg} \right\rangle
,\left| {ee} \right\rangle \}$,  if the initial state takes the form
\begin{eqnarray}
\label{rho}
\rho  = \left( {\begin{array}{*{20}c}
   a & 0 & 0 & e  \\
   0 & c & c & 0  \\
   0 & c & c & 0  \\
   e & 0 & 0 & b  \\
\end{array}} \right),
\end{eqnarray}
the form remains under evolution. This class of states includes the
easily prepared state as $\left| {gg} \right\rangle$ and $\left|
{ee} \right\rangle$. Since ${\rm Tr}\rho  = 1$, it can be found that $c =
(1 - a - b)/2$.

Concurrence \cite{Concurrence} is chosen to measure the
entanglement. For density matrix in Eq. (\ref{rho}),
\[
C(\rho ) = \max [0,1 - (\sqrt a  + \sqrt b )^2 ,2\left| e \right| + a + b - 1].
\]

Recently, a geometric measure of quantum discord (GMQD) is proposed \cite{discord3}, which is defined by
\[
D^g (\rho ) = \mathop {\min }\limits_{\chi  \in \Omega _0 } \left\| {\rho  - \chi } \right\|^2,
\]
where $\Omega_0$ denotes the set of zero-discord states and $\left\|
X \right\|^2  = {\rm Tr}X^2$ is the square norm in the Hilbert-Schmidt
space. For the density matrix Eq. (\ref{rho}), there is a simple
expression for its geometric measure of quantum discord:
\[
D^g (\rho ) = \min [D_1 ,D_2 ,D_3 ]/4,
\]
where $D_1  = ( - 1 + a + b + 2e)^2  + ( - 1 + a + b - 2e)^2$, $D_2
= (a - b)^2  + ( - 1 + 2a + 2b)^2  + ( - 1 + a + b - 2e)^2$, and
$D_3  = (a - b)^2  + ( - 1 + 2a + 2b)^2  + ( - 1 + a + b + 2e)^2$.

\section{Results}
In the following, we study the dynamics of the two atoms for different initial states.

\textit{Separable initial states}.---For the initial state $\left| {ee}
\right\rangle$, the solution is
\begin{eqnarray}
a(t) &=& \frac{e^{-2 t} }{6 (-2+\eta ) (-1+\eta )^2}[6 e^{2 t} (-1+\eta )^3-16 e^{t-\frac{t}{\eta }} (-2+\eta ) \eta ^2 \nonumber\\
& &+e^{\frac{4 t (-1+\eta )}{\eta }} (-3+\eta ) \eta  (1+\eta )+3 (-2+\eta ) (1+\eta ) (-1+3 \eta )],\nonumber \\
b(t)&=&e^{-2t},   \qquad  e(t)=\frac{2 e^{-2 t} \left(-1+e^{2 t+\frac{t (-1-\eta )}{\eta }}\right) \eta }{-1+\eta }. \nonumber
\end{eqnarray}
The concurrence increases in the first
period and then decreases (see FIG. 2 (a)). After it decreases to zero, it increases
again to steady value. More importantly, entanglement vanishes in finite time
when the detection is inefficiency, $\eta <1$. The duration of the
vanishing time increases as $\eta$ decreases. For the
steady concurrence, we can get a closed expression as $C(\rho (t \to \infty )) = 1/(2 - \eta )$. It is
an increasing function of $\eta$. When the efficiency of the
detection is perfect, the steady concurrence is $1$.

\begin{figure*}
\centering
\includegraphics[height=0.30\textheight,width=0.7\textwidth]{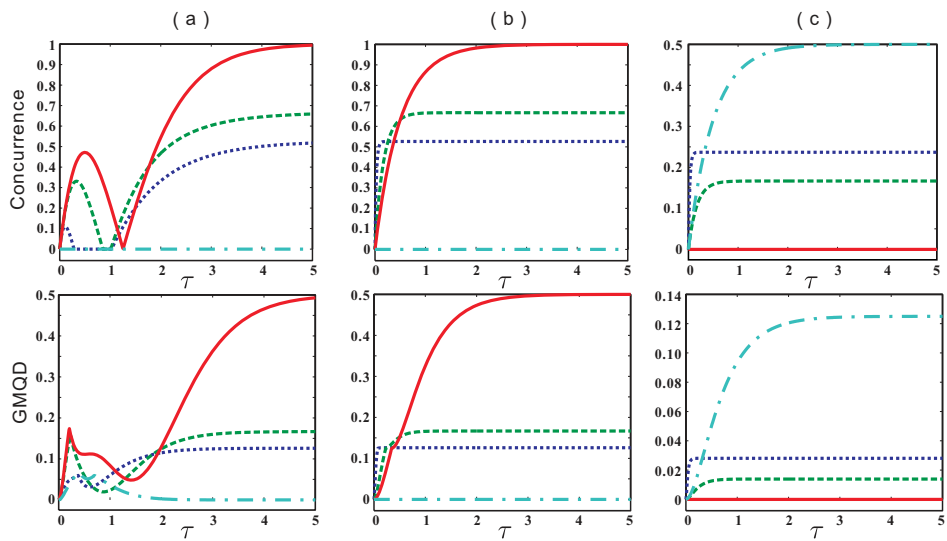}
\caption{\small (Color online) Time evolution of the concurrence and GMQD, with the initial state
being (a) $\left| {ee} \right\rangle$, (b) $\left| {gg} \right\rangle$, and (c) $\left| {eg} \right\rangle$,
all with $\lambda=1$, $\mu=1$,$\Omega=0$, for the case of  (i) $\eta=1$ (solid curve),
(ii) $\eta=0.5$ (dash curve), (iii) $\eta=0.1$ (dot curve), (iv) without feedback control (dot-dash curve).
 } \label{fig2}
 \end{figure*}

Since the density matrix Eq. (\ref{rho}) is specified by three parameters
($a$, $b$, and $e$), We can explain the finite-time vanishing of entanglement
induced by the inefficiency of the detection by the pictorial approach, similar to
\cite{pictorial}. In FIG. 3, separable states with the
form of Eq. (\ref{rho}) are in the region of the wedge shape. Initial state is located on
the point $(0, 1, 0)$. The final state is located on the $a$ axis.
Since the region of the valid states ($a, b \in [0, 1]$, $a+b\leq1$ and $\left| e \right|^{2} \leq a b$)
is separated by the region of separable states, only when the evolution of the state just crosses the
boundary of wedge shape the entanglement doesn't vanish in finite time. We can prove
that it only happens when $\eta=1$. For geometric measure of quantum
discord, it follows the same tendency as concurrence, but without
sudden death. It agrees with the general result that almost all states with none zero discord can never
lead to states with zero discord for a finite time interval for Markovian dynamics \cite{almost}.
From FIG. 2 (a) we can see that the geometric measure of quantum discord undergoes discontinuous change.
For the case there is no feedback, the dynamics of the discord is
different from concurrence. The concurrence remains zero all
the time, while the discord increases first and then goes to zero.
From FIG. 3, we can see that the evolution is confined in the region of separable states, but it
doesn't remain on the straight line $a+b=1, e=0$, where discord equals
zero, so the discord is not zero.
\begin{figure}
\centering
\includegraphics[height=0.40\textheight,width=0.4\textwidth]{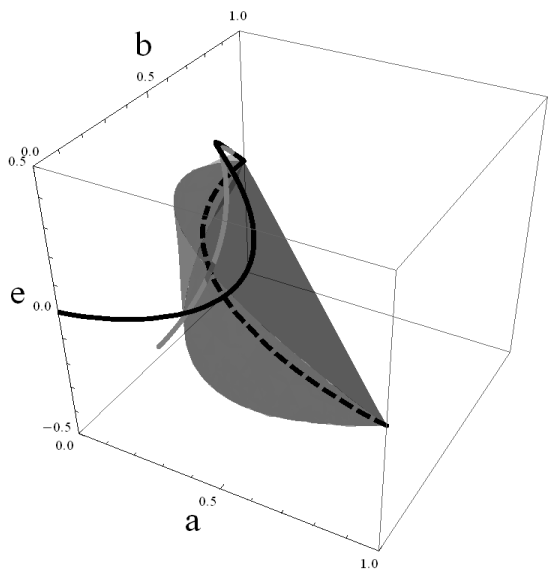}
\caption{\small (Color online) Dynamical evolution of the density matrix of two atoms with initial state $\left| {ee} \right\rangle$, for the case of (i) without feedback (Black dash curve), (ii) with perfect detection (Black solid curve) (iii)with imperfect detection $\eta=0.5$ (Gray solid curve). All the separable states are in the region of wedge shape.} \label{fig3}
\end{figure}

For the initial state $\left| {gg} \right\rangle$, the solution is given by:
\[
a(t)=\frac{-1-e^{\frac{2 t (-2+\eta )}{\eta }}+\eta }{-2+\eta }, \quad  b(t)=0, \quad e(t)=0.
\]

The concurrence and discord increases to the steady value
monotonely (see FIG. 2 (b)). The dynamics is very simple.
But the question is where the concurrence and discord come from.
From the initial condition, we can get that the concurrence and discord are not
from the initial state of the atoms, the environment, or the driving field.
The only source of the concurrence and discord is the control field. But since the the atom are
initially in the lower state, there is no photon emitted from the
atoms. So the detector can not be trigged by the photon from the atoms.
But this photon is not the unique source to trigger the detector.
The noise can also trigger the detector, too.
So we can say that the concurrence and discord is triggered by the noise.

For the initial state $\left| {eg} \right\rangle$, the form of the density matrix is not of the form of
Eq. (\ref{rho}). We solve the master equation by fourth-order Runge-Kutta
method. The concurrence and discord increase to steady values monotonely (see FIG. 2 (c)). But
the steady-state concurrence is inverse proportional to the
detection efficiency. That means the feedback is not always good for
getting a high concurrence and discord. More seriously, if the detection is
perfect, the concurrence and discord remains zero all the time.

\begin{figure*}
\centering
\includegraphics[height=0.30\textheight,width=0.9\textwidth]{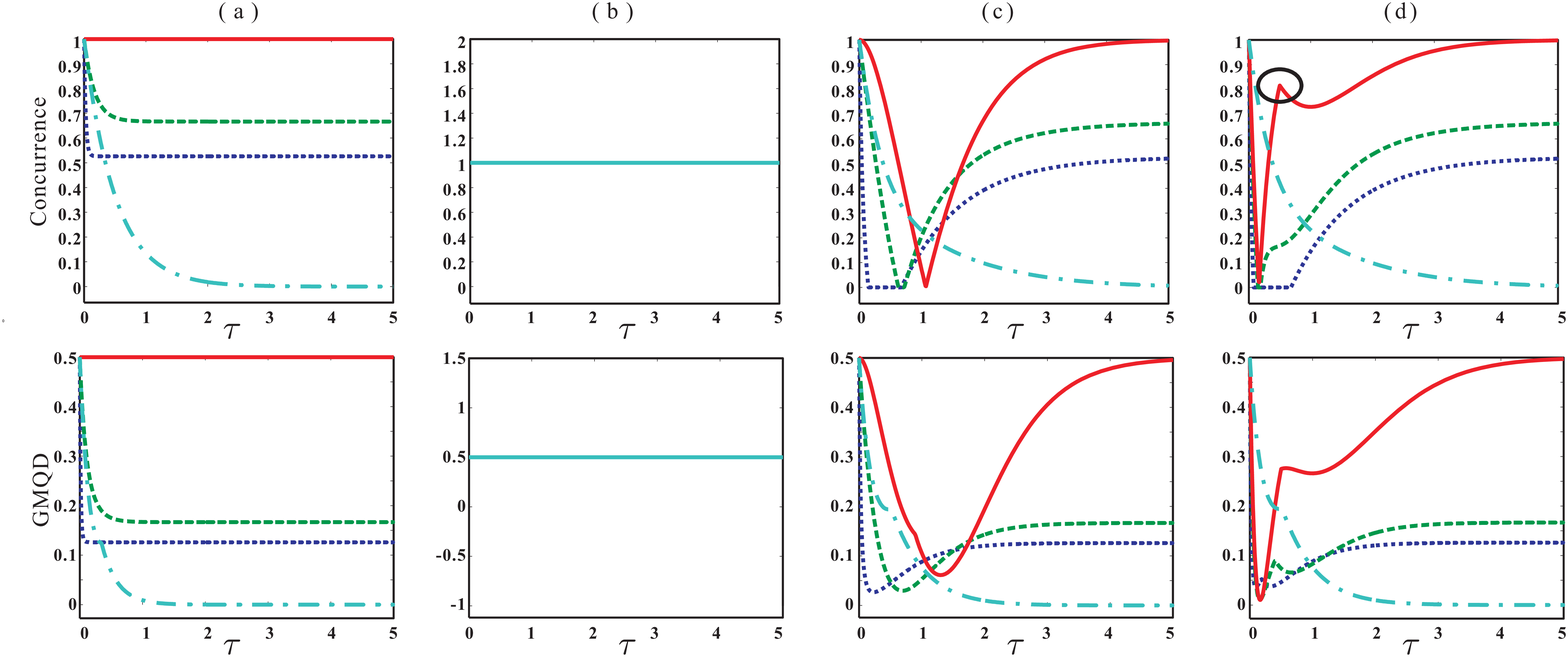}
\caption{\small (Color online) Time evolution of the concurrence and GMQD, with the initial state
being (a) $(\left| {eg} \right\rangle+\left| {ge} \right\rangle)/\sqrt{2}$,
(b) $(\left| {eg} \right\rangle - \left| {ge} \right\rangle)/\sqrt{2}$,
(c) $(\left| {ee} \right\rangle + \left| {gg} \right\rangle)/\sqrt{2}$,
and (d) $(\left| {ee} \right\rangle - \left| {gg} \right\rangle)/\sqrt{2}$,
all with $\lambda=1$, $\mu=1$,$\Omega=0$, for the case of (i) $\eta=1$ (solid curve),
(ii) $\eta=0.5$ (dash curve), (iii) $\eta=0.1$ (dot curve), (iv) without feedback control (dot-dash curve).} \label{fig4}
\end{figure*}

\textit{Entangled initial states}.---Dynamics for entangled initial states are also be
For initial state $\phi_{+}=(\left| {eg} \right\rangle +\left| {ge}
\right\rangle)/ \sqrt{2}$, the concurrence and discord decrease to
steady values (See FIG. 4 (a)). And the steady-state concurrence is
proportional to the detection efficiency. For the perfect detection,
the concurrence remains $1$ and the discord remains $1/2$ all the time.
In fact, detail analysis tells us that the state remains $\phi_{+}$.
Although $\phi_{+}$ is not a decoherence-free state under the
influence of the environment, it is a steady state for the
dynamics under the perfect feedback control \cite{feedbackentanglement2}.
We also study the case when the initial state is $\phi_{-}=(\left|
{eg} \right\rangle -\left| {ge} \right\rangle)/ \sqrt{2}$. Different from the case
of $\phi_{+}$, the concurrence remains $1$ and discord remains $1/2$ no matter
what $\eta$ is (See FIG. 4 (b)). That means the inefficiency of the detection doesn't
influence the dynamics.This is because, firstly, the initial state is in
the decoherence free subspace, so the state of the system doesn't
change under the influence of the environment, secondly, the $\phi_{-}$ is invariant under the control Hamiltonian, that means
even if the feedback control is triggered by the noise, the control
laser does not change the state of the system. So, we can say, the control
Hamiltonian is compatible with decoherence free subspace.

For the initial state $(\left| {ee} \right\rangle +\left|{gg}
\right\rangle)/ \sqrt{2}$, the concurrence and discord decrease at first, and
then increase to steady values (See FIG. 4 (c)). And when the
detection is imperfect, the concurrence vanishes in finite time.
For the initial state $(\left| {ee} \right\rangle -\left| {gg}
\right\rangle)/ \sqrt{2}$, the concurrence and discord decrease very
quickly, and then increase to steady values. And when the detection
is not perfect, the concurrence vanishes in finite time (See FIG. 4 (d)).
More importantly, for perfect detection, sudden change may also happen to concurrence,
at $t=0.5$. When $\eta=1$, there is a close expression of concurrence,
$C(\rho)={\rm Max}\left[0,C_1,C_2\right]$, with $C_1 = e^{ - 2t} [ - 1 + 2t - 2t^2  + e^{2t}  - \left| {1 - 2t} \right|]$
and $ C_2 = e^{ - 2t} \left| { - 1 + 2t} \right| - [1 - \frac{{e^{ - 2t} }}{2} - \frac{1}{2}e^{ - 2t} (1 - 2t)^2 ]$.
Sudden change happens at $t=0.5$. But it is not the time that the size relation of $C_1$ and
$C_2$ changes. Instead, at $t=0.5$, $C_1$ is not continuous. This
kind of sudden change is different from the sudden change of
discord. Although entanglement sudden death and revival can also be regarded as a kind of
sudden change happened when concurrence equals zero, in our case,
the sudden change happens at a nonzero value.
Since this kind of sudden change is very rare, this remains an open whether there is
some physical meaning behind it or it is just from the definition of concurrence.

\section{Conclusion}
To summarize, we study a model of two collectively damped atoms under practical feedback control. We focus
on the effect of the detection inefficiency on the dynamical and
asymptotic behavior of the entanglement and quantum discord. We find that
the inefficiency of detection induces the entanglement to vanish in finite time, and
can counterintuitively increase the asymptotic entanglement and quantum discord. The noise of detection
can trigger control field to create entanglement and quantum discord when no photon are emitted from the atoms.
More importantly, we find that the dynamics of entanglement also presents sudden change.

\section*{ACKNOWLEDGMENTS}
This work is supported by the Key Project of the National
Natural Science Foundation of China (Grant No. 60837004),
the National Hi-Tech Program of China (863 Program).

\end{document}